# A Survey of Brain Computer Interface Using Non-Invasive Methods

Ritam Ghosh
Department of Electrical Engineering
Vanderbilt University
ritam.ghosh@Vanderbilt.edu

## Abstract

**Research on Brain-Computer Interface (BCI) began in the 1970's and has increased in volume and diversified significantly since then. Today BCI is widely used for applications like assistive devices for physically challenged users, mental state monitoring, input devices for hands-free applications, marketing, education, security, games and entertainment. This article explores the advantages and disadvantages of invasive and non-invasive BCI technologies and focuses on use cases of several non-invasive technologies, namely electroencephalogram (EEG), functional Magnetic Resonance Imaging (fMRI), Near Infrared Spectroscopy (NIRs) and hybrid systems.**

## 1 Introduction

The brain is the largest and the most complex of all human organs. It consists of billions of nerve cells called neurons that send information to other nerve cells, muscles or gland cells. The brain has unmatched computational capacity and is capable of parallel processing information streams pouring in from multiple sense organs simultaneously and generating appropriate responses. It is capable of learning and memory, consciousness and emotions. Its extremely complex structure and exceptional computational capabilities have intrigued researchers from the early ages. Various paradigms like neuroscience, cognitive sciences, machine learning and artificial intelligence, brain computer interface etc. have been developed to understand the structure and functioning of the brain in more detail.

A brain-computer interface is defined as a communication system that does not depend on the brain's normal output pathways of peripheral nerves and muscles [1]. The main inspiration behind the development of BCI was to provide alternate means of control to people suffering from neuro-muscular disabilities. This led to the designing of several assistive devices that facilitated the restoring of lost motor abilities of physically challenged individuals [2]. Mobile robots [3,4] and BCI based prosthetic limbs also called neuro-prosthetic devices [5,6] are being used to help people regain normal functionality. Research has been conducted to develop methods to help people recover from stroke and various rehabilitation methods have been suggested. BCI based virtual reality setups have been used to gather data from stroke patients which was later used to control robotic prosthetics [7]. BCI has also been used to restore communication for people who are completely or partially paralyzed using different techniques ranging from yes/no binary capabilities [8] to virtual keyboards and spellers [9]. BCI's have also been used to evaluate the mental state of a subject for monitoring performance capability [10]. Other applications include workload monitoring, browsing and other media applications [11] and even as a sole or additional control input to games. The key to all these applications is in extracting reliable and meaningful information from the activities of the brain and devising methods and algorithms to extract features from it. Several methods and devices have been developed over time to 'read' the activities of the brain.

Neurons communicate with each other using electrical signals through physical connections or by exchanging chemicals called neurotransmitters. While communicating, the neurons exhibit an increase in the consumption of oxygen and glucose and hence cause an increase in blood flow to the active regions of the brain. Using various brain imaging technologies, it is possible to observe the electrical, chemical, or blood flow changes as the brain processes information or responds to various stimuli. The multichannel measurements from the instruments are then used to create a map of the activity patterns of the brain and from that we can infer the specific cognitive processes occurring in the brain at any given time. The different technologies used to accomplish this are discussed in the following section.

# 2   Types of Brain Imaging Technology: Advantages and Disadvantages

Different techniques have been developed to measure the activity levels of the different regions of the brain. The methods can be broadly classified into three categories: non-invasive, semi-invasive and invasive. In non-invasive BCI, the sensors are placed on the scalp over the skin. In semi-invasive BCI, the sensors are placed under the skin on the surface of the brain and in invasive BCI, the micro-electrodes are placed directly inside the cortex to measure the activity of single neurons [12].

Examples of non-invasive brain controlled interface methods include electroencephalography (EEG) where electrodes measure electric potentials produced by different regions of the brain, magnetoencephalography (MEG) where probes measure the magnetic fields generated by the currents in the brain, functional magnetic resonance imaging (fMRI) where sensors detect the change in magnetic characteristics of haemoglobin when it is oxygenated or deoxygenated, functional near infrared spectroscopy (fNIRS) which measure the hemodynamic responses associated with neural behavior by the difference in absorption characteristics of oxygenated and deoxygenated haemoglobin. Non-invasive techniques suffer from low spatial and/or temporal resolution and low signal to noise ratio. MRI and MEG require bulky and expensive equipment and hence experiments are confined to very limited lab setups. On the other hand, unlike the semi-invasive and invasive techniques, these methods do not require any surgical implants and hence the chance of medical complications do not exist. These are relatively cheaper and more convenient to use and safer for the subjects than the other two categories and hence most BCI research uses non-invasive techniques and so it benefits from wide availability of data and literature [13].

Semi-invasive techniques include electrocorticography (ECoG) which uses electrodes placed on the exposed surface of the skin to measure electrical activity from the cerebral cortex. Though ECoG and EEG record similar types of signals, ECoG exhibit higher special resolution and signal fidelity and is relatively resistant to noise. It has lower risk of medical complications than invasive methods and produces signals with higher amplitude than non-invasive methods. But it still requires a craniotomy to implant the electrodes and hence is performed only when a surgery is required for medical reasons. This makes it inconvenient for general use despite having superior signal recording ability [13].

Invasive techniques involve micro electrodes implanted directly into the brain during a neurosurgery. These can be used to collect data from a single area of the brain or multiples areas. The quality of the signal is the highest with the maximum resolution and signal to noise ratio. But with time, the body reacts to the presence of the electrodes by building scar tissues around them which over time reduce the resolution and SNR of the system. The surgery can be risky and hence is performed mainly on paralyzed subjects and often on blind subjects. In fact, one of the most significant applications of the invasive BCI technique was the restoration of partial eyesight to patients with acquired blindness developed by Dr. William Dobelle [14].

In the following sections, a few non-invasive methods of brain computer interface are described in detail.

# 3   Electroencephalography (EEG)

## 3.1   Data Acquisition Mechanism

It is the most commonly used and the oldest non-invasive technique for brain imaging. It operates by using metal electrodes placed on the scalp to measure the electrical activity of the neurons. A conductive electrolytic gel is applied on the electrodes to increase the conductivity between the electrodes and the skin. The position where the electrodes are placed is governed by some protocols to ensure consistency and repeatability. The most common system for placing the electrodes is the international 10-20 system which provides twenty-one standard positions. With the advent of higher resolution EEG devices with more electrodes, the international 10-10 system with eighty-one standard electrode placement positions was introduced and later the even more tightly packed 10-5 system with more than three hundred electrode placement positions was introduced.

Brain signals recorded using an EEG device are classified based on their frequency, called brain rhythmic activity or EEG rhythms [15]. The main frequencies of EEG signals in humans are:

1) Delta (<3 Hz): These waves have the highest amplitude and lowest frequency. It is the dominant rhythm in infants up to one year of age and in stages 3 and 4 of sleep. It may occur focally with subcortical lesions and in general distribution with diffuse lesions, metabolic encephalopathy hydrocephalus or deep midline lesions. It is usually most prominent frontally in adults (e.g. FIRDA - Frontal Intermittent Rhythmic Delta) and posteriorly in children e.g. OIRDA - Occipital Intermittent Rhythmic Delta).

2) Theta (3.5 Hz – 7.5 Hz): It is classified as the slow activity of the brain. It is typical in children up to thirteen years of age and in sleep but abnormal in awake adults. It can be seen as a manifestation of focal subcortical lesions. It can also be seen in generalized distribution in diffuse disorders such as metabolic encephalopathy or some instances of hydrocephalus.

3) Alpha (7.5 Hz – 13 Hz): It can be best observed in the posterior part of either side of the brain and is greater in amplitude on the dominant side. It is the dominant rhythm in normal relaxed humans above thirteen years of age but disappears if alerted or thinking,

4) Beta (> 14 Hz): It is known as the fast activity of the brain. It is usually seen on both sides in symmetrical distribution and is most evident frontally. It is accentuated by sedative-hypnotic drugs especially the benzodiazepines and the barbiturates. It may be absent or reduced in areas of cortical damage. It is generally regarded as a normal rhythm. It is the dominant rhythm in patients who are alert or anxious or have their eyes open. [16]

## 3.2  Data Processing

Once the raw data is collected, it is processed through a series of steps before being used for BCI systems. These steps involve pre-processing, feature selection and classification. One of the most common pre-processing methods is Independent Component Analysis (ICA) where the mixed signal can be decomposed into its statistically independent components. The raw signals are first centered i.e. the mean is subtracted from the raw data so that the resultant signal is an oscillatory signal centered around zero. Then this data is ready for feature extraction.

The temporal and frequency information in EEG data are characterized using hybrid features [17]. The common features used for EEG data are the coefficients of the auto-regressive model that is fit to the data, features related to the power spectral density and third order statistics like sum of logarithmic amplitudes. A method to extract features using the first layers of a multilayer perceptron has also been suggested [18] where once the whole network has been trained with data from each electrode as the input to a node in the input layer, the first hidden layer is examined and the nodes whose sum of weights is close to zero is assumed to have low discriminative power. These features are then used for classification using methods like Linear Discriminant Analysis (LDA), Support Vector Machines (SVM) etc.

## 3.3  BCI Systems and Applications

One of the most significant signals used in EEG based BCI is the P300 signal [19]. It is an event related potential generated as a result of decision making. In EEG it shows up as a positive voltage spike which appears when a person encounters a target object among a group of non-target objects and the spike appears roughly 300ms after the stimulus is presented from which it gets its name. The P300 is widely used in BCI for assistive devices like spellers. A matrix of letters is shown to the user and each row is highlighted one by one. Once the user selects a row, each column is highlighted one by one and thus the user can select the desired letter. A combination of ML techniques is also used for word predictions to streamline the process.

 One of the major challenges in BCI research is subject training. To solve this problem, an online single trial EEG based BCI was proposed to control hand holding and hand grasping through an interactive virtual reality environment [20]. The goal was to test if untrained subjects could achieve satisfactory results online without offline classifier training. It was found that online training with feedback performed at par with the traditional method of data collection without feedback and subsequent offline classifier training. Also, since brain signals vary significantly from person to person and also from day to day, the use of an adaptive probabilistic neural network was introduced that performed well in classification of time varying EEG signals. The experimental evaluation showed an accuracy of 75.4 % after only three minutes of online training and a peak accuracy of 84% after significant online training.

The high temporal resolution of EEG facilitates the development of real time controllers for mobile robots. In [21] EEG based BCI was used to control a robotic wheelchair capable of going left or right. In the experimental space a grid was drawn with a left goal and a right goal. The subject was required to navigate to the required goal from the starting point. The subjects were asked to think left and think right a hundred times and the data was collected. Fifty of them was used for pattern generation and the remaining was used for pattern validation. To remove the physiological artefacts a band pass filter was used with a passband of 0.53 Hz – 30 Hz. An iterative recursive training algorithm was used to train the classifier. The experiment was set up such that to accomplish the target the number of correct decisions required was three and the maximum wrong decisions allowed was one. If the decision taken by the BCI was random, then the chance of still accomplishing the target was 31.2 %. On performing the experiment ten times each for left goal and right goal on each of six subjects, an accuracy of 80% was obtained. This experiment showed the validity of a wheelchair with EEG based BCI being the sole controller.

Another notable application of EEG based BCI is its use as a game controller for games designed to help subjects improve concentration or relaxation. In [22] two games "Brain Chi" and "Dancing Robot" have been developed to help patients train for concentration and relaxation. In "Brain Chi" the size of the protective shield of the character depends on the concentration of the user i.e. the user has to protect the character with their 'brain power'. In "Dancing Robot" the speed of dancing depends on the concentration of the user. When the user concentrates positive points are added to their score and when the concentration level is detected as low, points are deducted. The same game can be converted as a relaxation game by just reversing the logic. The notable feature of this work is the use of fractal dimension feature to detect concentration level instead of the power of EEG signals. Higher fractal dimensions correspond to more complex and irregular signals and lower fractal dimensions correspond to more regular signals. Higuchi and Box-counting algorithms were used for the calculation of the fractal dimensions.

# 4 Functional Magnetic Resonance Imaging (fMRI)

## 4.1 Data Acquisition Mechanism

Another commonly used non-invasive technique for BCI applications is functional magnetic resonance imaging (fMRI). The device is equipped with a very powerful electromagnet with a field strength of typically 3 teslas i.e. about 50000 times the Earth's magnetic field. The magnetic field generated by this magnet affects the orientation of the atoms in the targets brain. Normally the atoms are randomly oriented and hence the net magnetic field of the atoms cancel out. The magnetic field from the MRI machine orients the atoms in a particular way such that their own fields add up to be significant enough to be measured. During any neural activity, there is a demand for oxygen and the response is an increase in blood flow towards the region. Oxygen is carried by haemoglobin in blood and haemoglobin is diamagnetic when oxygenated but paramagnetic when deoxygenated. This difference in magnetic property leads to a difference in the magnetic resonance signal and the amount of difference depends on the degree of oxygenation. FMRI records this difference and interprets brain activity. Unlike EEG, which measures rapid changes in the cortical activity of the brain, fMRI indirectly measures neural activity by measuring the oxygen flow in blood i.e. the Blood Oxygen Level Dependent (BOLD) signal. The fMRI uses Echo Planar Imaging technique (EPI) to acquire brain activity slice by slice and creates an activity map. The real-time fMRI-BCI's are important because of their magnetic field strength, good spatial and moderate temporal resolution (in the range of mm and seconds respectively), better echo time, and good magnetic field homogeneity. However, unlike EEG, the BOLD signal reflects vascular effects and only indirectly the neuronal signal so the hemodynamic response function introduces a physiological delay of 3 to 6 s before signal changes can be observed. Despite the fact that BOLD is an indirect measure, there is growing evidence for a strong correlation between the BOLD signal and electrical brain activity. Studies have characterized the relationship between localized increases in neuronal activity and the corresponding increase in BOLD making it possible to interpret positive functional responses in terms of neural changes [23]. FMRI creates a 3D volume of the subjects' brain consisting of voxels with one intensity value per voxel per scan. This 3D structure is compressed and flattened into a single line representation and several such volumes from a single session are joined together to create a 4D data structure which represents the entire data for the session. This data is then used for further processing.

## 4.2 Data Processing

The typical first step in pre-processing MRI data is slice timing correction [24]. Since the image is produced in slices, various parts of the image are taken at different time points. This is difficult to account for in later analysis and hence a timing correction is applied to bring all the slices of a single image to the same time point reference. This is done by interpolating the voxel values for the time points where that particular voxel was not scanned (this assumes that the intensity of the voxels changes smoothly in a continuous curve).

Another very common pre-processing step is head motion correction [24]. When the head moves relative to the scanner, the 3D coordinates for the scanner remains the same but a different neuron is now present in that coordinate. This results in a voxel representing a neuron that was represented by some other voxel in the past. These distortions are removed by a rigid body transform where the entire matrix is translated and rotated by the same amount to make the time series of each voxel smooth. The difference between the intensities of each voxel in two consecutive time points is considered as a cost function and the rotation and translation is done in an iterative manner with a goal of reducing the cost function.

The non-uniformities of the field produced by the scanner causes some distortion in the resultant images. To account for this distortion, a field map is created by acquiring multiple images with multiple echo times. Then the final image is rectified using the field map [24].

Another common filtering used in FMRI data is temporal filtering where unwanted frequencies which are of no interest are eliminated. The time series data is converted to the frequency domain by observing the repeating values of the intensities of the voxels (Fourier transform), unwanted frequencies are removed, and the data is converted back to the time series using the inverse Fourier transform.

A common noise removal method is spatial filtering, which means averaging the intensities of neighboring voxels to create a smooth image. The averaging is often done by convolution with a Gaussian kernel. This improves the signal to noise ratio but at the cost of a slight reduction in spatial resolution.

After filtering and removing any distortions, the data is ready for statistical analysis. A very common approach is called the Generalized Linear Model (GLM). But this model does not take into account the contribution of relationships between multiple voxels. To rectify this, a newer statistical model called the Multi Voxel Pattern Analysis (MVPA) is used. The data is then used for training neural networks and other classifiers according to the application.

## 4.3  BCI Systems and Applications

One of the applications of fMRI based BCI is in emotional processing due to its high spatial resolution. In [25], fMRI based BCI was used to study the voluntary control of the anterior cingulated cortex (ACC) on emotional processing. There are two major subdivisions of the ACC: the dorsal ACC, also called the "cognitive division" (ACcD) and the rostral-ventral ACC, called the "affective division" (ACaD). Due to its involvement in different functional networks, physiological self-regulation was applied to study cognitive and emotional parameters, for example, emotional valence and arousal, dependent on the differential activation of the two subdivisions. In this study, two continuously updated curves were presented to the subject depicting BOLD activity in ACcd and ACad. During blocks of 60-second duration, subjects were instructed to move both curves upwards (alternating 60 seconds rest and 60 seconds up-regulation). The subject was instructed to use his own strategy for voluntary BOLD regulation. The subject reported the use of mental imagery of winter landscapes, engaging in snowboarding and social interaction as his strategy to increase the BOLD signal. During baseline blocks, he attended to the signal time-course without performing any specific task. The subject rated the valence of his affective state significantly more positive for activation blocks as compared to baseline blocks.

Study of neuroplasticity and functional reorganization for recovery after neurological diseases such as stroke is another area of research for BCI's. Real-time fMRI feedback has been successfully used to successively reactivate affected regions of the brain. In [26], the researchers trained four healthy volunteers to control the BOLD response of the supplementary motor area (SMA). Offline analysis showed significant activation of the SMA with training. Further, with training there was a distinct reduction in activation in the surrounding areas, indicating that volitional control training focuses activity in the region-of-interest.

Another important domain in which fMRI based BCI is extremely effective is in treating chronic pain. Chronic pain can be substantially affected by cognitive and emotional processes. Subregions within the rostral ACC in association with other brain regions are implicated to be involved in the perception of pain. Hence, it is possible that by altering the activity in the rostral ACC, pain perception might be accordingly varied. In [27], researchers reported a substantial decrease of symptoms in chronic pain patients by training patients to self-regulate ACC. A further report from the same group [28], involving 16 healthy volunteers and 12 chronic pain patients, indicates the potential application of real-time fMRI for treating chronic pain. Subjects were able to learn to control activation in the rostral anterior cingulate cortex (rACC), a region involved in pain perception and regulation. The authors reported that if subjects deliberately induced increases or decreases of rACC fMRI activation, there was a corresponding change in the perception of pain caused by an applied noxious thermal stimulus. Control experiments showed that this effect was not observed after training without real-time fMRI feedback, or using feedback from a different region, or sham feedback derived from a different subject. Chronic pain patients were also trained to control activation in rACC and reported decreases in the ongoing level of chronic pain after training.

# 5  Near Infrared Spectroscopy (fNIRS)

## 5.1  Data Acquisition Mechanism

The use of Near Infrared Spectroscopy (NIRs) in BCI applications is fairly recent. NIRS also measures the changes in blood flow as fMRI, but using a different technique, infrared light (650nm – 1000nm) instead of magnetic field to measure the concentration changes of oxygenated hemoglobin (HbO) and deoxygenated hemoglobin (HbR). NIRS measures the blood flow changes in the local capillary network caused by neuron firings. Since the hemoglobin is an oxygen carrier, the changes of HbO and HbR concentration levels after a neuronal activation can be related to the relevant neuronal firings. fNIRS uses near-infrared light emitter-detector pairs operating with two or more wavelengths. The light emitted into the scalp diffuses through the brain tissues resulting in multiple scattering of photons. Some of these photons exit the head after being partially absorbed and partially reflected by the cortical region of the brain, wherein the chromophores (i.e., HbO and HbR) are changing in time. These exited photons are then detected by using strategically positioned detectors. Since HbO and HbR have different absorption coefficients for different wavelengths of near infrared light, the relationship between the exiting-photon intensity and the incident-photon intensity can be used to calculate the changes of the concentrations of HbO and HbR along the path of the photons by applying the modified Beer-Lamberts law [29]. The distance between the emitter and detector determines the depth at which the measurement is taken, an increase in emitter-detector distance corresponds to an increase in imaging depth. To measure hemodynamic response signals from the cortical areas, an emitter-detector separation of around 3 cm was suggested in [30]. A separation of less than 1 cm might contain only skin-layer contribution, whereas that of more than 5cm might result in weak and therefore unusable signals. The number of 'channels' of measurement depend on the number of both emitters and detectors as well as the configuration of their placement relative to each other. A 'channel' of measurement is located at the mid-point of the straight line joining an emitter and detector and depth of the channel depends on the distance between the emitter and the detector e.g. a configuration of ten detectors arranged in an array of two rows and five columns and four emitters each placed at the center of a 'square' formed by the detectors gives rise to sixteen channels. fNIRs typically have better spatial resolution than EEG and better temporal resolution than fMRI which makes it a good choice for monitoring activities in the motor cortex and prefrontal cortex regions.

## 5.2  Data Processing

The acquired fNIRS signals contain various types of noise, which can be categorized into instrumental noise, experimental error, and physiological noise. Since the instrumental noise and experimental error are not related to the brain activities, it is better to remove them prior to converting the raw optical density signals to the concentration changes of HbO and HbR through the modified Beer-Lambert law.

Instrumental noise can be caused due to imperfections in the hardware design and assembly, limitations of the analog to digital convertors or due to external factors like other electronic gadgets in the vicinity. This type of noise is usually a constant signal of high frequency and can be removed using a low pass filter with a 3-5 Hz cutoff frequency. Experimental error is typically caused by head motion which causes the optodes to move away from their designated positions. This causes a sudden sharp change in the light intensity resulting in a spike in the resultant waveform. Several methods of motion artifact filtering are used like the Wiener filtering-based method [31], eigenvector-based spatial filtering [32] etc. Physiological noise include heartbeat (1 ∼ 1.5 Hz), respiratory noise (0.2 ∼ 0.5 Hz) and Mayer Waves (∼0.1 Hz), which are related to changes in blood pressure. These noises are removed by using a bandpass filter with typically 0.1 Hz - 0.4 Hz cut-off frequency, though filters with other cut-off frequencies have also been employed. Other methods like adaptive filtering, principal component analysis (PCA) and Independent Component Analysis (ICA) have also been used to remove them.

After filtering, the different brain activities are classified based on several features. Some features can be extracted from the raw intensity data while most are extracted from the hemodynamic signals (HbO, HbR, total haemoglobin HbT = HbO + HbR and cerebral oxygen exchange COE = HbO – HbR) which are computed from the raw data using the modified Beer-Lamberts law. Typical features used for NIRs based BCI are peak amplitude, mean, variance, slope, skew, kurtosis and zero-crossing. Some studies have also proposed the use of features like filter coefficients from Kalman filtering, recursive least squares estimation and wavelet transform for classification purposes. Some of the common classification techniques used are Linear discriminant analysis (LDA), support vector machines (SVM) and also more recently artificial neural networks (ANN).

### 5.3 BCI Systems and Applications

BCI systems have been shown to be effective in restoring some of the lost motor and/or cognitive functions in individuals with stroke and spinal cord injury. The underlying idea of doing so is the ability of BCI feedback to induce self-regulation of brain activity. EEG has the limitations of imprecise localization and inaccessibility of subcortical areas and fMRI, though more effective, typically involves bulky and expensive machines making them inconvenient. This makes fNIRS, with its good spatial resolution and decent temporal resolution, a very attractive option, in comparison with fMRI, in accessing subcortical brain signals. It is low cost, easy to use, and most of all it is portable. It can be used even in an ambulance. Moreover, fNIRS is less sensitive to motion artifacts because it can be fashioned into a cap or headband which can be securely attached to the subjects' head. This makes the potential of use of fNIRS in neurofeedback studies very high. In [33], the researchers demonstrated the possibility of using fNIRS-based neurofeedback to allow the users to willfully regulate their hemodynamic responses. In [34], studies revealed that fNIRS-based neurofeedback can be used for long-term training as well, and such repetitive neurofeedback can induce specific and focused brain activation, while in contrast, sham feedback (synthesized or from another subject) has led to diffuse brain activation patterns over broader brain areas.

Another important application of BCI is to serve as a means of communication for people with motor disorders such as Amyotrophic Lateral Sclerosis (ALS), spinal cord injury or Locked-In syndrome. In [35] researchers developed an fNIRS-BCI system for binary communication based on activations from the prefrontal area. The subjects were required to perform a specific task such as mental arithmetic or music imagery to increase the cognitive load and, thereby, respond "yes" or to remain relax and, thus, respond "no" to a given question. The average accuracies obtained with online classification were approximately 82%. In [36], researchers proposed an fNIRS-BCI-based online word speller. Their system involves using right-hand and left-hand motor imagery to move a cursor on a two-dimensional to select letters.

Another important application of fNIRS-BCI is the restoration of movement capability for people with motor disabilities. The control commands generated by a BCI system can be used to control a prosthetic limb or a wheelchair. It is desirable to have a portable system for these applications so that the user can move freely. These applications, for safety purposes, cannot afford high error rates, and must be fast enough to provide real-time control. The portability and resilience to motion artefacts and moderate temporal resolution make fNIRs a good choice for this type of application. Several fNIRS-BCI studies [37] have tried to improve classification accuracies and information transfer rates which will make these applications more feasible.

## 6 Hybrid BCI Systems

Each BCI type has its own shortcoming and disadvantages. To utilize the advantages of different types of BCIs, different approaches are combined, called hybrid BCIs [38]. In a hybrid BCI, two or more different types of BCI systems are combined. It is also possible to combine one BCI system with another system which is not BCI-based, for example, combining a BCI system with an electromyogram (EMG)-based system. Different techniques and their combinations are utilized based on the application that the hybrid BCI is going to be used for. The main purpose of combining different systems to form a "hybrid" BCI is to improve accuracy, reduce errors, and overcome disadvantages of each conventional BCI system. In general, in a hybrid BCI, two systems can be combined sequentially or simultaneously [39]. In a simultaneous hybrid BCI, both systems are processed in parallel. Input signals used in simultaneous hybrid BCIs can be two different brain signals or one brain signal and another input. In sequential hybrid BCIs, the output of one system is used as the input of the other system. This approach is mostly used when the first system task is to indicate that the user is trying to communicate to the system, sort of a switch for the second system.

A combination of two types of brain signals can be collected using the same device to increase accuracy, reliability and compatibility for a wide range of users. In [40], a hybrid system utilizing Steady State Visually Evoked Potential (SSVEP) and Event Related Desynchronization (ERD) was proposed using EEG data. The proposed hybrid was evaluated during a task that compared ERD based BCI, SSVEP based BCI, and ERD-SSVEP based hybrid BCI. During the ERD BCI task, two arrows appeared on the screen. When the left arrow appeared, subjects were instructed to imagine opening and closing their left hand and for the right arrow, subjects imagined opening and closing the right hand. In the SSVEP task, subjects were instructed to gaze at either left (8 Hz) or right (13 Hz) LED depending on which cue appeared. In the hybrid task, when the left arrow was showed, subjects were imagining the left hand opening and closing while gazing at the left LED simultaneously. The task was similar for the right arrow. Results show the average accuracy of 74.8% for ERD, 76.9% for SSVEP, and 81.0% for hybrid.

An example of two types of brain signals being used sequentially can be found in [41] where the very reliable P-300 signal and SSVEP have been used. The P300 and SSVEP combination work well as the stimuli for evoking both patterns can be shown on one screen simultaneously. The P300 paradigm considered in this study is a 6×6 speller matrix based on the original P300 row/column paradigm. Only one frequency is allocated for the SSVEP paradigm. The background color was flashed with a frequency slightly less than 18 Hz. The background color change facilitates the SSVEP detection. During the classification, P300 and SSVEP signals were separated by a band pass filter. The SSVEP was utilized as a control state (CS) detection. When the user was gazing at the screen, the SSVEP was detected and it was assumed that the user intended to send a command. At that point the P300 based speller system was activated. For SSVEP detection, the mean power spectral density (PSD) in the narrow band near the desired frequency and the PSD in the wider range near the desired frequency were utilized in an objective function. These values were subtracted from each other and divided over the PSD value from the wide band and the function value was compared to a specified threshold.

Another possible combination for a hybrid BCI is P300 and motor imagery (MI)- based BCI. The basic concept in this type of hybrid is based on the features of P300 and ERD/ERS in control applications. P300 is a reliable BCI type for selecting one item out of several items and can be used for discrete control commands. On the other hand, due to the low degree of freedom presented by MI-based BCI, this type of BCI is more efficient for continuous control commands. These two types of BCIs can be joined to present more complicated control commands in one task. In [42], a P300-Motor Imagery Hybrid BCI for controlling a wheelchair in a home environment is introduced. The wheelchair control commands were divided into three steps: destination selection, navigation and stopping. For destination selection the user had to select the destination of the wheelchair motion by selecting one of the items among a list of destinations. To implement this control command, an accurate and reliable interface is needed, and false acceptance rates should be as low as possible. For this task, a P300 BCI presented at a screen was utilized. The experiments on healthy subjects showed a response time of about 20 seconds, the false acceptance rate 2.5% and the error less than 3%. The results showed that P300 was an appropriate option for the interface. For the navigation of the wheelchair, an autonomous motion control was introduced. The destination was selected, and the wheelchair started its motion toward the destination following virtual guiding paths. A proximity sensor was considered for stopping the wheelchair facing obstacles. For the stop command, the interface needs to be fast, reliable and have a low false acceptance rate. Two approaches for a stopping command were presented. The first approach was the fast P300, in which, on the screen, there is only one item "The Stop" and the task is the detection of user's intention. Experimental results showed reduction in response time. However, increase in false acceptance makes this approach inapplicable. The second approach was to use a mu-beta BCI. The position of a cursor was considered for presenting the visual feedback for the mu-beta BCI system and the control of the cursor was based on an arm movement imagination. Results showed approximately the same response time as the fast P300 approach, but for false acceptance, a rate of zero was achieved. Since the low false acceptance rate and fast response are the most important needs for this type of BCI, the mu-beta BCI was considered a more reliable system for this application.

Another very effective hybrid BCI system can be composed by using two types of brain imaging devices simultaneously to improve reliability. In [43], EEG and NIRS measurements were utilized simultaneously for ERD-based BCIs. In this study, the experiment consisted of 2 blocks of motor execution and 2 blocks of motor imagery. For all blocks, both EEG and NIRS were measured simultaneously. The increase in concentration of oxygenated hemoglobin (HbO) and decrease in concentration of deoxygenated hemoglobin (HbR) were measured using NIRS. The global peak cross-validation accuracy for each subject was considered for evaluation of the hybrid BCI. The mean classification accuracies of HbO, HbR, and EEG separately for executed movement tasks were 71.1%, 73.3%, and 90.8%. For motor imagery tasks they were 71.7%, 65.0%, and 78.2% respectively. The mean classification accuracies of EEG/HbO, EEG/HbR, and EEG/HbO/HbR for executed movement tasks were 92.6%, 93.2%, and 87.4%, and for motor imagery tasks were 83.2%, 80.6%, and 83.1%, respectively. It was shown that the combination of EEG and NIRS improved the classification accuracy in both MI and executed movement tasks.

# 7 Conclusions

In this article, different types of brain-controlled interfaces were reviewed, and their working principles, advantages, disadvantages and use-cases were highlighted. A brain-computer interface is defined as a communication system that does not depend on the brain's normal output pathways of peripheral nerves and muscles. When the workload of any region of the brain increases due to various types of activities, certain observable changes occur. These changes may be in the form of electrical pulses, chemical changes, change in blood flow and change in the oxygen level of blood. Several devices have been developed to monitor these changes using a variety of physical principles. These devices are known as brain imaging devices. They can be invasive, semi-invasive or non-invasive. This article discusses the advantages and disadvantages of each type and elaborates on different types of non-invasive techniques to monitor the activities of the brain. Using these devices to facilitate the communication of the user with a computing system to monitor, enhance or enable certain activities is the core focus of the subject known as BCI. Emphasis is placed on the non-invasive techniques because these are convenient, can be used outside of a very controlled clinical facility and much more accessible and affordable than their invasive and semi-invasive counterparts. These are the qualities that have enabled the widespread use of such devices in research and commercial applications and thus have generated a rich collection of data and literature. Each type of imaging device and data collection technique has its own disadvantages and shortcomings, this led to the concept of hybrid BCI techniques which use different imaging devices or different modalities from the same device to improve reliability and accuracy and also increase the functionality of the system. Research has shown that the operation of the BCI depends on two adaptive controllers: the human brain, which gets better with practice at certain activities like motor imagery, and the brain imaging device controller, which gets better as more data is generated and it is better tuned to recognize certain activities with more precision and reliability. Successful implementations of BCI require the user to learn and maintain new skills which are not normally required in daily life. The subjects can be trained in these new skills using real time biofeedback from the BCI. This observation has led to separate line of research on the ways in which BCI can be used for the rehabilitation of patients of sclerosis, stroke and other motor disorders. This idea has also led to research on the ways BCI can be used to augment and enhance the communication capabilities of typical humans with computing devices to increase efficiency and provide more 'channels' of control input. BCI consists of several components like the raw data collecting device, filters, feature extraction, classification and user training tools. This makes BCI development a truly interdisciplinary endeavor combining physical principles, statistics, data science, electrical engineering, signal processing, algorithm design, psychology, neuroscience and knowledge of rehabilitation techniques.